\documentclass[12pt]{iopart}
\usepackage{amsfonts}
\usepackage{amsbsy}

\begin{document}

\renewcommand\vec[1]{\boldsymbol{#1}}
\newcommand{\myfrac}[2]{\frac{\displaystyle #1}{\displaystyle #2}}
\newcommand{\sn}{\mathop{\mathrm{sn}}\nolimits}
\newcommand{\cn}{\mathop{\mathrm{cn}}\nolimits}
\newcommand{\dn}{\mathop{\mathrm{dn}}\nolimits}

\title{KdV-Volterra chain.}
\author{G.M. Pritula$^{1}$ and V.E. Vekslerchik$^{1,2}$}

\address{$^{1}$
Institute for Radiophysics and Electronics, Kharkov, Ukraine
}

\address{$^{2}$
Universidad de Castilla-La Mancha, Ciudad Real, Spain
}

\ead{galinapritula@yandex.ru, vadym.vekslerchik@uclm.es}

\begin{abstract}
This paper is devoted to the system of coupled KdV-like equations. 
It is shown that this apparently non-integrable system possesses an 
integrable reduction which is closely related to the Volterra chain. 
This fact is used to construct the hyperelliptic solutions of the 
original system.
\end{abstract}

\submitto{\JPA}

\maketitle

\section{Introduction.}

The equation we are going to study is
\begin{equation}
  \left( \partial_{t} - \partial_{xxx} \right) q_{n} = 
  6 \left( q_{n+1} q_{n}^{3} q_{n-1} \right)_{x}.  
\label{kdvc-eq}
\end{equation}
The integrability/non-integrability of this apparently new system is now an 
open question. 
On one side, in this paper we establish its relationship 
with the well-known integrable system, the Volterra hierarchy (VH) 
\cite{M1974,KvM1975}
(that is why we will call (\ref{kdvc-eq}) the 'KdV-Volterra chain' (KdVVC)), 
and present the $N$-phase periodic solutions from which 
one can derive the $N$-soliton ones.
According to the widely used hypothesis the existence of more than two-soliton 
solutions is a strong evidence of the integrability of the problem.
On the other side, the system (\ref{kdvc-eq}) can be easily reduced 
by taking $q_{n}=q$ to the generalized KdV equation,
\begin{equation}
  q_{t} - q_{xxx} = 30 \, q^{m} q_{x}  
\end{equation}
with $m=4$, that is not of solitonic type for $m>2$ \cite{Miura1976}. 
This statement does not have a rigorous proof and is based on the 
Painlev\'{e}-like tests and numerical studies of interaction between solitary 
waves (see e.g. \cite{AC1992}). However, after so many years it seems 
improbable that the integrability of the generalized KdV equation can be 
established. Consequently, according to the Ablowitz-Ramani-Segur hypothesis 
(another integrability test that is not proved but has a long history of 
succesfull applications) the KdVVC is non-integrable. 
Here we have an apparent contradiction between the 
more than two-soliton and Ablowitz-Ramani-Segur tests: the former indicates 
the integrability of the KdVVC while the latter leads to the opposite 
conclusion. Probably we have a situation of the conditional integrability 
introduced by Dorizzi \textit{ et al} \cite{DGRW1986} 
when equations in high dimensions (the higher KP equations in the case 
of \cite{DGRW1986}) are not individually integrable but become such, if one 
demands that their solutions also satisfy the lower equations of the 
hierarchy. As one can see below, our situation is almost the same.
So, it is clear that the question of integrability/non-integrability of 
(\ref{kdvc-eq}) is not trivial and deserves a special study, which is out of 
the scope of this paper. The main result of the presented work is the fact 
that equation (\ref{kdvc-eq}) is one of a small number of nonlinear 
$(2+1)$-dimensional systems for which an infinite family of explicit 
solutions can be derived.

In this paper after reducing equation (\ref{kdvc-eq}) in section \ref{sec-VH} 
to VH (namely this reduction is the main topic of this paper) 
and discussing the hyperelliptic solutions for the latter 
(section \ref{sec-qps}) we present the corresponding solutions for 
the KdVVC (sections \ref{sec-sol} and \ref{sec-red}).

\section{Reduction to the VH. \label{sec-VH}}

The main result of this paper can be formulated as follows: 
if functions  $\tau_{n}$ solve the system 
\begin{equation}
  \left\{
  \begin{array}{l}
  D_{x} \, \tau_{n} \cdot \tau_{n-1} = \tau_{n+1} \tau_{n-2}  
  \\
  \left( D_{t} - D_{xxx} \right) \tau_{n} \cdot \tau_{n-1} 
  = 3 \tau_{n+2} \tau_{n-3} - 6a \tau_{n+1} \tau_{n-2} 
  \end{array}
  \right.
\label{volt-syst}
\end{equation}
where $a$ is a constant, $D_{xxx}=D_{x}^{3}$ and $D_{t}$ and $D_{x}$ 
are Hirota's bilinear operators,
\begin{equation}
  D_{t}^{m}D_{x}^{m}  \, u \cdot v = 
  \left.
  \left( \frac{ \partial }{ \partial t' } \right)^{m} 
  \left( \frac{ \partial }{ \partial x' } \right)^{n} \;
  u\left( t+t',x+x' \right) \, v\left( t-t', x-x' \right) 
  \right|_{t'=x'=0}, 
\end{equation}
then the quantities 
\begin{equation}
  p_{n} = \frac{ \tau_{n+1} \tau_{n-1} }{ \tau_{n}^{2} }
\label{p-def}
\end{equation}
solve the equation 
\begin{equation}
  \left( \partial_{t} - c \, \partial_{x} - \partial_{xxx} \right) p_{n} = 
  6 \left( p_{n+1} p_{n}^{3} p_{n-1} \right)_{x}  
\label{kdvc-eq-mod}  
\end{equation}
with some function $c=c(x,t)$. In the case of constant $c$ (\ref{kdvc-eq-mod}) 
is related to (\ref{kdvc-eq}) by simple Galilean transformation:
$
  q_{n}(t,x) = p_{n}(t,x-ct).
$

To prove this statement let us start from the first equation of system 
(\ref{volt-syst}) which we rewrite as
\begin{equation}
  \partial_{x} \mu_{n} = 
  \frac{ \tau_{n+1}\tau_{n-2} }{ \tau_{n}\tau_{n-1} } 
\label{d-mu}
\end{equation}
where
\begin{equation}
  \mu_{n} = \ln\frac{ \tau_{n} }{ \tau_{n-1} }. 
\end{equation}
From (\ref{d-mu}) we can obtain the following equations for the functions 
$p_{n}$ (\ref{p-def}) that are related to $\mu_{n}$ by 
$p_{n} = \exp\left( \mu_{n+1} - \mu_{n} \right)$:
\begin{eqnarray}
  \partial_{x} p_{n} & = & 
  p_{n} \left( u_{n+1} - u_{n} \right)
\label{dp-dx}
\\
  \partial_{xx} p_{n} & = & 
  p_{n} \left( 
    w_{n+1} - 4w_{n} + w_{n-1} 
    + u_{n+1}^{2} + u_{n}^{2}
  \right)
\label{dp-dxx}
\end{eqnarray}
where, in order to make the formulae more readable, we have introduced
\begin{eqnarray}
  u_{n} & = & p_{n} p_{n-1}
  =
  \frac{ \tau_{n+1}\tau_{n-2} }{ \tau_{n}\tau_{n-1} } 
  \\[3mm]
  w_{n} & = & u_{n+1} u_{n} = p_{n+1} p_{n}^{2} p_{n-1} 
  = 
  \frac{ \tau_{n+2} \tau_{n-2} }{ \tau_{n}^{2} }.
\end{eqnarray}
On the other hand, differentiating (\ref{d-mu}) we obtain
\begin{equation}
  \partial_{xx} \, \mu_{n} = w_{n} - w_{n-1}
\end{equation}
which implies
\begin{equation}
  \partial_{xx} \ln\tau_{n} = w_{n} + b
\label{ddln}
\end{equation}
where $b$, in general, is a function of $t$ and $x$ but does not depend on 
the index $n$. 

Now we can calculate
$ D_{xxx} \; \tau_{n} \cdot \tau_{n-1} $
and then  $\partial_{t} \mu_{n}$ and $\partial_{t} p_{n}$.
Starting from the well-known formula for the Hirota's operators,
\begin{equation}
  \frac{ D_{xxx} \, a \cdot b  }{ ab } =
  \Lambda_{xxx} 
  + \Lambda_{x}^{3} 
  + 3 \Lambda_{x} \left( \ln ab \right)_{xx},
  \qquad
  \Lambda = \ln\frac{ a }{ b },
\end{equation}
after a little lengthy but simple calculations we get 
\begin{eqnarray}
  \frac{ D_{xxx} \; \tau_{n} \cdot \tau_{n-1} }{ \tau_{n} \tau_{n-1} } 
  & = & 
  u_{n} \left( 
      w_{n+1} + 2w_{n} + 2w_{n-1} + w_{n-2}  
  \right.
\\&&
  \left.\qquad
      + u_{n+1}^{2} + u_{n}^{2} + u_{n-1}^{2} - 2 u_{n+1}u_{n-1} 
      + 6b
  \right). 
\nonumber
\end{eqnarray}
Substituting this result into the second equation of (\ref{volt-syst}) 
we come to 
\begin{eqnarray}
  \left( \partial_{t} - c \, \partial_{x} \right) \mu_{n} 
  & = & 
  u_{n} \left( 
      w_{n+1} + 2w_{n} + 2w_{n-1} + w_{n-2}  
  \right.
\label{dmu-dt}
\\&&
  \left.\qquad
      + u_{n+1}^{2} + u_{n}^{2} + u_{n-1}^{2} + u_{n+1}u_{n-1}  
  \right) 
\nonumber
\end{eqnarray}
with $c=6(b-a)$,
from which one can derive the expression for $\partial_{t} \ln p_{n}$.
However, before doing that it seems useful to present (\ref{dmu-dt}) 
by means of the formulae for the derivatives of $u_{n}$ and $w_{n}$ 
stemming from (\ref{dp-dx})
\begin{eqnarray}
  \partial_{x} \, u_{n} & = & 
  u_{n} \left( u_{n+1} - u_{n-1} \right)
\\
  \partial_{x} \, w_{n} & = & 
  w_{n} \left( u_{n+2} + u_{n+1} - u_{n} - u_{n-1} \right)
\end{eqnarray}
in the following two forms:
\begin{eqnarray}
  \left( \partial_{t} - c \, \partial_{x} \right) \mu_{n} & = & 
  \phantom{-} \left( \partial_{x} + u_{n} - u_{n-1} \right) A_{n} + B_{n-1} 
\label{dmu-dt-one}
\\
  & = & 
  - \left( \partial_{x} + u_{n+1} - u_{n} \right) A_{n} + B_{n} 
\label{dmu-dt-two}
\end{eqnarray}
with
\begin{eqnarray}
  A_{n} & = & u_{n} \left( u_{n+1} + u_{n} + u_{n-1} \right) 
\\
  B_{n} & = & 
  u_{n+1}u_{n} \left( 2u_{n+2} + 3u_{n+1} + 3u_{n} + 2u_{n-1} \right).
\end{eqnarray}
Using (\ref{dmu-dt-one}) to calculate 
$\left( \partial_{t} - c \, \partial_{x} \right) \mu_{n+1}$
and (\ref{dmu-dt-two}) to calculate 
$\left( \partial_{t} - c \, \partial_{x} \right) \mu_{n}$
we can easily obtain 
\begin{eqnarray}
	\left( \partial_{t} - c \, \partial_{x} \right) \ln p_{n} & = & 
	\left( \partial_{t} - c \, \partial_{x} \right) \mu_{n+1} 
	- 
	\left( \partial_{t} - c \, \partial_{x} \right) \mu_{n} 
\\ & = & 
  \left( \partial_{x} + u_{n+1} - u_{n} \right) 
  \left( A_{n+1} + A_{n} \right)
\end{eqnarray}
or, recalling the fact that $u_{n+1} - u_{n-1} = \partial_{x} \ln p_{n}$,
\begin{equation}
	\left( \partial_{t} - c \, \partial_{x} \right) p_{n} =  
  \partial_{x} \, p_{n} \left( A_{n+1} + A_{n} \right).
\end{equation}
Comparing the above expression with (\ref{dp-dxx}), 
\begin{equation}
  p_{n} \left( A_{n+1} + A_{n} \right) = 
  \partial_{xx} \, p_{n} + 6 w_{n} p_{n},
\end{equation}
we obtain
\begin{equation}
  \left( \partial_{t} - c \, \partial_{x} - \partial_{xxx} \right) p_{n} =
  6 \, \partial_{x} \left( w_{n} p_{n} \right) 
\end{equation}
which is nothing but (\ref{kdvc-eq-mod}).

Equations (\ref{volt-syst}) belong to the VH, one of the classical integrable 
models for which a wide range of solutions has already been constructed  such 
as, e.g., solitons \cite{M1974,KvM1975} and quasiperiodic solutions
\cite{V1988,BGHT1998,T2000,GMHT2008}.
The main idea behind the present work is to use these results (with minor 
modifications) to derive solutions for (\ref{kdvc-eq}) that is possible due 
to the established relation between (\ref{volt-syst}) and (\ref{kdvc-eq}). 
To illustrate this approach we obtain the periodic solutions for (\ref{kdvc-eq}) 
starting from the hyperelliptic ones for the VH obtained 
in \cite{BGHT1998} (see also \cite{V1988} for the one-phase (elliptic) case). 
In the next section we rederive these solutions using an alternative 
to the methods in \cite{BGHT1998,T2000,GMHT2008} and present the formulae we 
need for our purposes.

\section{\label{sec-qps} Hyperelliptic solutions for the VH.}

Consider the hyperelliptic Riemann surface $\Gamma$
\begin{equation}
  \Gamma: \qquad
  w^{2} = \mathcal{P}_{2g+1}(\xi) = 
  \prod_{i=1}^{2g+1} \left( \xi - \xi_{i} \right)
\end{equation}
which is a compact Riemann surface of the genus $g$.
One can choose a set of closed contours (cycles) $\{ a_{i}, b_{i} \}_{i=1,
..., g}$ with the intersection indices
\begin{equation}
a_{i} \circ a_{j} =
b_{i} \circ b_{j} = 0,
\qquad
a_{i} \circ b_{j} = \delta_{ij}
\qquad
i,j = 1, \dots, g
\end{equation}
and find $g$ independent holomorphic differentials 
$\omega_{k}$ 
satisfying the normalization conditions
\begin{equation}
  \oint_{a_{i}} \omega_{k} = \delta_{ik},
  \qquad
  i,k = 1, \dots, g.
\end{equation}
The matrix of the $b$-periods,
\begin{equation}
\Omega_{ik} = \oint_{b_{i}} \omega_{k}
\end{equation}
determines the so-called period lattice,
$L_{\Omega} = \left\{
   \vec{m} + \Omega \vec{n},
   \quad
   \vec{m}, \vec{n} \in {\mathbb Z}^{g}
   \right\}$,
the Jacobian of this surface
$\mathrm{Jac}(\Gamma)={\mathbb C}^{g}/L_{\Omega}$ (2$g$ torus) and the Abel
mapping $\Gamma \to \mathrm{Jac}(\Gamma)$,
\begin{equation}
  P \to \int^{P}_{P_{0}} \vec{\omega}
\label{abel}
\end{equation}
where $P$ is a point of $\Gamma$,
$ P = \left( w, \xi \right) $,
$\vec{\omega}$ is the $g$-vector of the 1-forms,
$\vec{\omega} =
\left( \omega_{1}, \dots, \omega_{g} \right)^{\scriptscriptstyle T}$, 
and $P_{0}$ is some fixed point of $\Gamma$.

A central object of the theory of the compact Riemann surfaces is the
$\theta$-function, $\theta(\vec{\zeta})=\theta(\vec{\zeta},\Omega)$,
\begin{equation}
\theta\left(\vec{\zeta}\right) =
\sum_{ \vec{n} \, \in \, \mathop{\mathbb Z}\nolimits^{g} }
\exp\left\{
     \pi i \, \biggl( \vec{n}, \Omega \vec{n} \biggr) \; +
   2 \pi i \, \biggl( \vec{n}, \vec{\zeta} \biggr)
\right\}
\end{equation}
where $( \vec{n},\vec{\zeta} )$ stands for 
$\sum_{i=1}^{g} n_{i}\zeta_{i}$, 
which is a quasiperiodic function on $\mathbb{C}^{g}$
\begin{eqnarray}
\theta\left(\vec{\zeta} + \vec{n}\right) &=&
   \theta\left(\vec{\zeta}\right)
\\
\theta\left(\vec{\zeta} + \Omega\vec{n}\right) &=&
   \exp\left\{
      -   \pi i \, \biggl( \vec{n}, \Omega \vec{n} \biggr) \;
      - 2 \pi i \, \biggl( \vec{n}, \vec{\zeta} \biggr)
   \right\}
\theta\left(\vec{\zeta}\right)
\end{eqnarray}
for any $\vec{n} \in {\mathbb Z}^{g}$.

The calculations presented below are based on the famous Fay's trisecant formula 
\cite{Fay,Mumford2} that can be written as

\begin{equation}
  \varepsilon^{P_{4}}_{P_{3}} \, 
  \varepsilon^{P_{2}}_{P_{1}} \; 
  \theta \, 
  \theta^{P_{3}P_{4}}_{P_{1}P_{2}} 
  - 
  \varepsilon^{P_{3}}_{P_{1}} \, 
  \varepsilon^{P_{4}}_{P_{2}} \; 
  \theta^{P_{3}}_{P_{2}} \, 
  \theta^{P_{4}}_{P_{1}} 
  + 
  \varepsilon^{P_{3}}_{P_{2}} \, 
  \varepsilon^{P_{4}}_{P_{1}} \; 
  \theta^{P_{3}}_{P_{1}} \, 
  \theta^{P_{4}}_{P_{2}} 
  = 0. 
\label{fay}
\end{equation}
Here
\begin{equation}
  \theta^{Q_{1}...Q_{m}}_{P_{1}...P_{m}} = 
  \theta\left( 
    \vec{\zeta} 
    + \sum_{i=1}^{m}\int\nolimits^{Q_{i}}_{P_{i}} \vec{\omega} 
    \right) 
\end{equation}
and the skew-symmetric function $\varepsilon^{Q}_{P}$, 
$\varepsilon^{Q}_{P}=-\varepsilon^{P}_{Q}$, 
is closely related to the prime form \cite{Mumford2} and is given by
\begin{equation}
  \varepsilon^{Q}_{P} = 
  \theta\left( \vec{e} + \int\nolimits^{Q}_{P} \vec{\omega} \right)
\end{equation}
where $\vec{e}$ is a zero of the $\theta$-function: $\theta\left(\vec{e}\right)=0$.

In what follows we will fix the points $P_{1}$, $P_{2}$ and $P_{3}$ as the 
images of the points $\xi=0$ and $\xi=\infty$,
\begin{equation}
  \begin{array}{lcl}
  P_{1} & = & O = \left( + w_{0}, 0 \right)
  \\
  P_{2} & = & \infty 
  \\
  P_{3} & = & \overline{O} = \left( - w_{0}, 0 \right)
  \end{array}
\qquad
  w_{0} = \sqrt{\mathcal{P}_{2g+1}(0)}
\label{ABC}
\end{equation}
and will consider the fourth point appearing in (\ref{fay}) as a variable one. 
Definition (\ref{ABC}) leads to the possibility 
to take such integration paths in (\ref{fay}) that give
\begin{equation}
  \int\limits_{O}^{\infty} \vec{\omega} 
  +
  \int\limits_{\overline{O}}^{\infty} \vec{\omega} 
  = \vec{0}
\end{equation}
(to this end it is enough, for example, to connect the points $O$ and 
$\overline{O}$ with $\infty$ by the curves in $\Gamma$ having the same 
projections on the $\xi$-plane).

Now, by elementary transformations, we can convert the \textit{bilinear} 
identity (\ref{fay}) into the form which gives us solutions for the VH.

Making the shift 
$ \vec{\zeta} \to \vec{\zeta}_{n} = \vec{\zeta} + n \vec{\nu}$ and   
introducing the function $\Theta_{n}(P)$ by 
\begin{equation}
  \Theta_{n}(P) = \theta\left( \vec{\zeta}_{n} + \vec{\delta}(P) \right)
\end{equation}
where the vectors $\vec{\nu}$ and $\vec{\delta}(P)$ are given by
\begin{equation}
  \vec{\nu} = 
  \int\limits_{\overline{O}}^{\infty} \vec{\omega}, 
  \qquad 
  \vec{\delta}(P) = 
  \int\limits_{O}^{P} \vec{\omega}
\end{equation}
the Fay's identity (\ref{fay}) can be rewritten as
\begin{equation}
  u(P) \, \Theta_{n-1}(O) \, \Theta_{n}(P) 
  - 
  \Theta_{n}(O) \, \Theta_{n-1}(P) 
  = 
  \tilde{u}(P) \, \Theta_{n-2}(O) \, \Theta_{n+1}(P) 
\label{Fay-Theta}
\end{equation}
with
\begin{equation}
  u(P) = 
  \frac{ \varepsilon^{\overline{O}}_{O} }{ \varepsilon^{\infty}_{O} }, 
  \,
  \frac{ \varepsilon^{P}_{\infty} }{ \varepsilon^{P}_{\overline{O}} } 
  \qquad
  \tilde{u}(P) = 
  \frac{ \varepsilon^{P}_{O} }{ \varepsilon^{P}_{\overline{O}} }. 
\end{equation}
This equation implies that the function
\begin{equation}
  T_{n}(P) = \alpha^{ n^{2}/2 } [u(P)]^{n} \Theta_{n}(P) 
\label{T-def}
\end{equation}
where $\alpha$ is a constant that will be defined later solves 
\begin{equation}
  T_{n-1}(O) \, T_{n}(P)
  - 
  T_{n}(O) \, T_{n-1}(P)
  =  
  \tilde\xi_{\alpha}(P) \; 
  T_{n-2}(O) \, T_{n+1}(P)
\end{equation}
with
\begin{equation}
  \tilde\xi_{\alpha}(P) = 
  \frac{ 1 }{ \alpha^{2} } \, 
  \left(
    \frac{ \varepsilon^{\infty}_{O} }{ \varepsilon^{\overline{O}}_{O} } 
  \right)^{2} 
  \frac{ \varepsilon^{P}_{O} \, \varepsilon^{P}_{\overline{O}} } 
       { \left( \varepsilon^{P}_{\infty} \right)^{2} }.
\end{equation}
Noting that the fraction
$ \varepsilon^{P}_{O} \, \varepsilon^{P}_{\overline{O}} 
\left/ \left( \varepsilon^{P}_{\infty} \right)^{2} \right.$
has the same zeroes and poles as the projection 
$\xi(P)$,
\begin{equation}
  \xi(P): \qquad
  P = \left(w,\xi\right) \to \xi
\end{equation}
one can conclude that it is possible to find the value of the constant $\alpha$ 
which ensures
\begin{equation}
  \tilde\xi_{\alpha}(P) = \xi(P). 
\end{equation}

Now we have to introduce the dependence of $\vec{\zeta}$ on an infinite 
number of `times' $t_{j}$ in such a way that the shift
\begin{equation}
  \vec{\zeta} \to \vec{\zeta} + \vec{\delta}(P) 
\end{equation}
is the Miwa's shift
\begin{equation}
  \vec{\zeta}\left( \mathrm{t} \right) = 
  \vec{\zeta}\left( ..., t_{j}, ... \right)
  \to 
  \vec{\zeta}\left( \mathrm{t} + [\xi] \right) = 
  \vec{\zeta}\left( ..., t_{j} + \frac{\xi^{j}}{j}, ... \right).
\end{equation}
To do this, let us take $\xi$ as a local parameter near the point 
$O$ of $\Gamma$. Thus the forms $\omega_{i}$ can be presented as 
\begin{equation}
  \omega_{i} = \tilde\omega_{i}(\xi) \, d\xi
  = \sum_{k=0}^{\infty}\tilde\omega_{ik} \, \xi^{k} d\xi
\end{equation}
which leads to
\begin{equation}
  \int_{O}^{P} \omega_{i} = 
  \int_{0}^{\xi} \tilde\omega_{i}(\eta) d\eta
  = \sum_{k=0}^{\infty} \tilde\omega_{ik} \frac{ \xi^{k+1} }{ k+1 }
\end{equation}
and
\begin{equation}
  \vec{\delta}(P) = 
  \sum_{j=1}^{\infty} \vec{\zeta}_{j} \frac{ \xi^{j} }{ j }
\label{delta-j-def}
\end{equation}
where $\vec{\zeta}_{j}$ is the vector with the components $\tilde\omega_{i,j-1}$. 
Now it is easy to check that if we take 
\begin{equation}
  \vec{\zeta}\left( \mathrm{t} \right) = 
  \sum_{j=1}^{\infty} \vec{\zeta}_{j} \, t_{j}, 
\label{zeta-def}
\end{equation}
then
\begin{equation}
  \vec{\zeta}\left( \mathrm{t} + [\xi] \right) - 
  \vec{\zeta}\left( \mathrm{t} \right) = 
  \vec{\delta}(P).
\end{equation}
In a similar way using the expansion 
\begin{equation}
  \ln u(P) = 
  \sum_{j} \lambda_{j} \xi^{j} 
\label{lambda-j-def}
\end{equation}
one can introduce the function
\begin{equation}
  \phi\left( \mathrm{t} \right) =  
  \sum_{j} \phi_{j} t_{j} 
\label{phi-def}
\end{equation}
with
\begin{equation}
  \phi_{j} = j \lambda_{j} 
\end{equation}
such that
\begin{equation}
  u(P) = 
  e^{ \phi\left( \mathrm{t} + [\xi] \right) - \phi\left( \mathrm{t} \right) }. 
\label{u-phi} 
\end{equation}
Combining the above results one can present the functions $T_{n}(P)$ as 
functions $\tau_{n}\left( \mathrm{t} \right)$:
\begin{equation}
  T_{n}(O) = \tau_{n}\left( \mathrm{t} \right),
  \qquad
  T_{n}(P) = 
  \tau_{n}\left( \mathrm{t} + [\xi] \right)
\end{equation}
where
\begin{equation}
  \tau_{n}\left( \mathrm{t} \right) = 
  \alpha^{ n^{2}/2 } e^{ n\phi\left( \mathrm{t} \right)} 
  \theta\left( \strut \vec{\zeta}\left( \mathrm{t} \right) + n \vec{\nu} \right) 
\label{tau-qps}
\end{equation}
and conclude that functions $\tau_{n}\left( \mathrm{t} \right)$
solve 
\begin{equation}
  \tau_{n-1}\left( \mathrm{t} \right) \, 
  \tau_{n}\left( \mathrm{t} + [\xi] \right) 
  - 
  \tau_{n}\left( \mathrm{t} \right) \, 
  \tau_{n-1}\left( \mathrm{t} + [\xi] \right) 
  =  
  \xi \; 
  \tau_{n-2}\left( \mathrm{t} \right) \, 
  \tau_{n+1}\left( \mathrm{t} + [\xi] \right).
\label{vh-fr}
\end{equation}
Equation (\ref{vh-fr}) is nothing but the so-called functional representation 
of the VH \cite{V2005}. Expanding it in the power series in $\xi$ 
one can show that any of its solutions also solves equations 
of the VH, the first three of which are given by
\begin{eqnarray}
  && D_{t_{1}} \, \tau_{n} \cdot \tau_{n-1} = 
  \tau_{n+1} \tau_{n-2}
\label{vh-bilin-a}
\\[2mm] &&
  D_{t_{2}} \, \tau_{n} \cdot \tau_{n-1} = 
  D_{t_{1}} \, \tau_{n+1} \cdot \tau_{n-2}
\\[2mm] &&
  \left( 8D_{t_{3}} + D_{t_{1}t_{1}t_{1}} \right) \, \tau_{n} \cdot \tau_{n-1} = 
  \left( 6D_{t_{2}} + 3D_{t_{1}t_{1}} \right) \tau_{n+1} \cdot \tau_{n-2}
\end{eqnarray}
Eliminating the $t_{2}$-derivatives one can rewrite the last equation as
\begin{equation}
  \left( D_{t_{3}} - D_{t_{1}t_{1}t_{1}} \right) \, \tau_{n} \cdot \tau_{n-1} = 
  3 \tau_{n+2} \tau_{n-3}
  - 6b \, \tau_{n+1} \tau_{n-2}
\label{vh-kdvc}
\end{equation}
where, recall, 
$ b = \partial_{t_{1}t_{1}}\ln\tau_{n} - w_{n}$
and does not depend on $n$ by virtue of (\ref{vh-bilin-a}).
Thus the functions $\tau_{n}$ given by (\ref{tau-qps}) together with 
(\ref{zeta-def}) and (\ref{lambda-j-def}), (\ref{phi-def}) 
are hyperelliptic solutions for equations (\ref{volt-syst}) with $a=b$. 
This means that we have everything necessary to construct the 
hyperelliptic solutions for the KdVVC (\ref{kdvc-eq}).

\section{\label{sec-sol} Hyperelliptic solutions for the KdVVC.}

As it can be seen from the results of the previous section, the 
$\tau$-functions (\ref{tau-qps}) constructed from the hyperelliptic 
$\theta$-functions satisfy equations (\ref{volt-syst}), (\ref{ddln}) 
with $a=b$. This means that we do not need to know the exact values of 
the constants $a$ and $b$ because the parameter $c$ appearing in  
(\ref{kdvc-eq-mod}) for $p_{n}$ is automatically equal to zero in 
this case. In other words, we do not need to make the Galilean 
transformation and the hyperelliptic solutions for the KdVVC, $q_{n}$, 
are given by 
\begin{equation}
  q_{n} = \frac{ \tau_{n+1} \tau_{n-1} }{ \tau_{n}^{2} }.
\label{q-qps}
\end{equation}

Recalling the definition of $\tau_{n}$ (\ref{tau-qps}) one can note that 
the functions $\phi\left( \mathrm{t} \right)$ cancel themselves due to the 
structure of (\ref{q-qps}). 
We have introduced the function $u(P)$ in (\ref{T-def}), 
and hence the function $\phi\left( \mathrm{t}\right)$ in (\ref{u-phi}), 
to rewrite the Fay's identity as the standard bilinear 
functional equations of the VH \cite{V2005}. 
In principle one can deduce directly from 
(\ref{Fay-Theta}) some functional equations for the quantities
$\Theta_{n-1}\Theta_{n+1} / \Theta_{n}^{2}$ which lead to the differential 
equations solved by $p_{n}$ and $q_{n}$. However, in this case one has to 
perform more cumbersome calculations giving up the advatages of the bilinear 
approach. Moreover, we have not used this way because one of the goals of this 
work is to demostrate the relationship of (\ref{kdvc-eq}) with the standard 
Volterra equations.
The disappearance of the factor 
$e^{n\phi\left( \mathrm{t} \right)}$
from the final formulae for $q_{n}$ is a typical effect of the bilinear and 
the inverse scattering approaches: periodic solutions of integrable equations 
lead to quasiperiodic solutions of corresponding linear and bilinear problems. 
In some sense, the Fay-like formulae of the previous section contain more 
information than we need for the present work: for example, the above mentioned 
function $\phi\left( \mathrm{t} \right)$ which is inessential for $q_{n}$ would 
be crucial if we were deriving the Baker-Akhiezer function related to 
(\ref{q-qps}). 

The role of the parameter $\alpha$ is different: the factor 
$\alpha^{ n^{2}/2 }$ 'survives' the bili\-nearization procedure of calculating 
$\tau_{n-1}\tau_{n+1} / \tau_{n}^{2}$ and appears in the final formulae as the 
'amplitude' of the solutions.

Thus,
\begin{equation}
  q_{n} = 
  \alpha
  \frac{ 
    \theta\left(\vec{\zeta}_{n+1} \right) 
    \theta\left(\vec{\zeta}_{n-1} \right) }
  { \theta^{2}\left(\vec{\zeta}_{n} \right) }, 
  \qquad
  \vec{\zeta}_{n} = \vec{\zeta} + n\vec{\nu}.
\label{q-qps-struct}
\end{equation}
The dependence of $q_{n}$ on $x$ and $t$ can be established by comparing 
(\ref{volt-syst}) and (\ref{vh-kdvc}). It is obvious that we have to 
identify $x$ with $t_{1}$ and $t$ with $t_{3}$ neglecting all the other 
times $t_{k}$, $k \ne 1,3$ (they play the role of a constant):
\begin{equation}
  \vec{\zeta}(x,t) = 
  \vec{\zeta}_{1} \, x + \vec{\zeta}_{3} \, t + \vec{\zeta}_{*},
  \qquad
  \vec{\zeta}_{*} = \mbox{constant}
\label{q-qps-zeta}
\end{equation}
where $\vec{\zeta}_{1,3}$ are the vectors with the components
\begin{equation}
  \begin{array}{lclcl}
  \left( \vec{\zeta}_{1} \right)_{i} & = & 
  \tilde\omega_{i0} & = & \tilde\omega_{i}(0)
  \\[2mm]
  \left( \vec{\zeta}_{3} \right)_{i} & = & 
  \tilde\omega_{i2} & = & \frac{1}{2} \tilde\omega_{i}''(0).
  \end{array}
\end{equation}

\section{\label{sec-red} Elliptic solutions for the KdVVC.}

In this section we rewrite solutions (\ref{q-qps-struct}), 
(\ref{q-qps-zeta}) and calculate all the constant parameters in terms of 
the Jacobi elliptic functions. 
Expression (\ref{q-qps-struct}) for $g=1$ becomes
\begin{equation}
  q_{n} = 
  \alpha
  \frac{ 
    \vartheta_{3}\left( \zeta_{n+1} \right) 
    \vartheta_{3}\left( \zeta_{n-1} \right) 
  }
  { \vartheta_{3}^{2}\left( \zeta_{n} \right) },
  \qquad
  \zeta_{n} = \zeta + n\nu
\end{equation}
(we use the notation of the book \cite{Bateman}) 
and can be rewritten using the Jacobi functions as 
\begin{equation}
  q_{n} = q_{*} f\left( z_{n} \right) 
\end{equation}
where
\begin{equation}
  f(z) = f(z; a, k) = 1 - k^{2} \sn^{2}(a,k) \sn^{2}(z, k ) 
\label{ell-sol-f}
\end{equation}
and
\begin{equation}
  z_{n} =  K(k) \left( 2\zeta_{n} + 1 \right),
  \qquad
  a =  2 K(k) \nu  
\end{equation}
with $K(k)$ being the complete elliptic integral of the first kind.
Starting from the standard formulae for the derivatives of the elliptic 
functions one can obtain the following identity for the function $f(z)$:
\begin{equation}
  f''(z) = \beta_{2} f^{2}(z) + \beta_{1} f(z) + \beta_{0}
\label{f-prop-beta}
\end{equation}
where
\begin{equation}
  \begin{array}{lcl}
  \beta_{2} & = & 
  - \myfrac{ 6 }{ \sn^{2} a }
  \\[4mm]
  \beta_{1} & = & 
    \myfrac{ 12 }{ \sn^{2} a } 
  - 4 \left( 1 + k^{2} \right)
  \\[4mm]
  \beta_{0} & = & 
  - \myfrac{ 6 }{ \sn^{2} a } 
  + 4 \left( 1 + k^{2} \right)
  - 2 k^{2} \sn^{2} a. 
  \end{array}
\end{equation}
Another useful property of the function $f(z)$ stems from the 
superposition formulae for the elliptic functions:
\begin{equation}
  f(z+a) f^{2}(z) f(z-a) = \gamma_{1}f(z) + \gamma_{0} 
\label{f-prop-gamma}
\end{equation}
where
\begin{equation}
  \begin{array}{lcl}
  \gamma_{1} & = & 4 \cn^{2} a \dn^{2} a
  \\[2mm]
  \gamma_{0} & = & f^{2}(a) - 4 \cn^{2} a \dn^{2} a.
  \end{array}
\end{equation}
Using (\ref{f-prop-beta}) and (\ref{f-prop-gamma}) one can verify the 
fact that the functions
\begin{equation}
  q_{n}(x,t) = q_{*} f\left( z + na \right)  
\label{ell-sol-struct}
\end{equation}
with
\begin{equation}
  z = z(x,t) = \kappa x + \varpi t + \mbox{constant}.
\label{ell-sol-zeta}
\end{equation}
satisfy the equation
\begin{equation}
  \kappa^{3} \frac{ d^{2} q_{n} }{ dz^{2} } 
  - \varpi q_{n} 
  + 6\kappa q_{n-1}q_{n}^{2}q_{n+1}
  = C = \mbox{constant}
\end{equation}
provided
\begin{equation}
  \kappa^{2} = - 6\frac{ \gamma_{1} }{ \beta_{2} }q_{*}^{4}, 
  \qquad
  \varpi = \beta_{1} \kappa^{3} + 6 \gamma_{0} \kappa q_{*}^{4} 
\end{equation}
(the constant $C$ is given by $C=\gamma_{0} \kappa^{3} q_{*}$) 
which implies that functions $q_{n}$ solve (\ref{kdvc-eq}). 
Substituting the values of $\beta_{i}$, $\gamma_{i}$ we come to the 
final result: the elliptic solutions of the KdVVC are given by 
(\ref{ell-sol-struct}), (\ref{ell-sol-f}) and (\ref{ell-sol-zeta}) 
with
\begin{eqnarray}
  \kappa & = & 2 \sn a \cn a \dn a \; q_{*}^{2}
  \\[2mm]
  \varpi & = & 
  6 \left[ 
      \frac{ 1 }{ \sn^{2} 2a } 
    + \frac{ 1 }{ \sn^{2} a } 
    - \frac{2}{3} \left( 1 + k^{2} \right)
  \right] \, \kappa^{3}.
\end{eqnarray}

\section*{Acknowledgements.}

This work has been partially supported by grants FIS2007-29093-E  
(Ministerio de Educaci\'on y Ciencia, Spain) 
and PCI-08-0093 (Consejer\'ia de Educaci\'on y Ciencia, Junta de Comunidades
de Castilla-La Mancha, Spain).

\section*{References}                                                          %


\begin{thebibliography}{99}

\bibitem{M1974}
  S.V. Manakov, 1974,
  Complete integrability and stochastization of discrete dynamical systems.
  \textit{Soviet Phys. JETP} \textbf{40}, 269--274.  
\bibitem{KvM1975}
   M. Kac and P. van Moerbeke, 1975, 
   On an explicitly soluble system of nonlinear differential equations 
   related to certain Toda lattices. 
   \textit{Advances in Mathematics}, \textbf{16}, 160--169.
\bibitem{Miura1976}
   R.M. Miura, 1976, 
   The Korteweg-de Vries equation: A survey of results.
   \textit{SIAM Review}, \textbf{18}, 412--459.
\bibitem{AC1992}
   M.J. Ablowitz and P. A. Clarkson, 1992,
   \textit{Solitons, nonlinear evolution equations and inverse scattering.}
   (Cambridge: Cambridge University Press) 
\bibitem{DGRW1986}
   B. Dorizzi, B. Grammaticos, A. Ramani, P. Winternitz, 1986, 
   Are all the equations of the Kadomtsev-Petviashvili hierarchy integrable?
   \textit{J. Math. Phys.}, \textbf{27}, 2848-2852.
\bibitem{V1988}
   V.L. Vereshchagin, 1988,
   Hamiltonian structure of averaged difference systems.
   \textit{Matematicheskie Zametki}, \textbf{44}, 584-–595,
   English translation: \textit{Mathematical Notes}, \textbf{44}, 798--805.
\bibitem{BGHT1998}
   W. Bulla, F. Gesztesy, H. Holden,  and G. Teschl, 1998,
   \textit{Algebro-geometric quasi-periodic finite-gap solutions of the Toda 
   and Kac-van Moerbeke hierarchies.}
   Memoirs of the American Mathematical Society, \textbf{135} 
   (Providence: American Mathematical Society).
\bibitem{T2000}
   G. Teschl, 2000,
   \textit{Jacobi operators and completely integrable nonlinear lattices.}
   Mathematical Surveys and Monographs, \textbf{72}, 
   (Providence: American Mathematical Society).
\bibitem{GMHT2008}
   F. Gesztesy, J. Michor, H. Holden, G. Teschl, 2008,
   \textit{Soliton equations and their algebro-geometric solutions. 
   II: (1+1)-dimensional discrete models.}
   (Cambridge: Cambridge University Press)
\bibitem{Fay}
  J. Fay, 1973,
  \textit{Theta functions on Riemann surfaces.}
  Lect. Notes in Math., {\bf 352} (Berlin, Heidelberg: Springer).
\bibitem{Mumford2}
  D.Mumford, 1984, {\it Tata lectures on Theta II.}
  (Boston: Birkhauser).
\bibitem{V2005}
  V.E. Vekslerchik, 2005, 
  Functional representation of the Volterra hierarchy. 
  {\it J. of Nonlin. Math. Phys.}, {\bf 12}, 409--431.
\bibitem{Bateman}
  H. Bateman, 1955,
  \textit{Higher Transcendental Functions}, Volume 3.
  Based, in part, on notes left by Harry Bateman
  and compiled by the staff of the Bateman manuscript project, 
  director Arthur Erd\'{e}lyi
  (New York, Toronto, London: Mc Graw-Hill).
\end{thebibliography}
\end{document}